\documentclass[pdflatex,iicol,sn-mathphys-num]{sn-jnl}
\usepackage{graphicx}%
\usepackage{multirow}%
\usepackage{tabularx}%
\usepackage{amsmath,amssymb,amsfonts}%
\usepackage{amsthm}%
\usepackage{mathrsfs}%
\usepackage[title]{appendix}%
\usepackage{textcomp}%
\usepackage{manyfoot}%
\usepackage{booktabs}%
\usepackage{algorithm}%
\usepackage{algorithmicx}%
\usepackage{algpseudocode}%
\usepackage{listings}%
\usepackage{dcolumn}
\usepackage{bm}
\usepackage{enumitem}
\usepackage{breqn}

\newcolumntype{C}{>{\centering\arraybackslash}X}

\begin{document}
	\title{Formation of cylindrical shells via sphere packing from fluidized beds\\
	\normalsize{\textcolor{blue}{An edited version of this paper was published by Springer in Open Access: Oliveira, V.P.S., Borges, D.S., Franklin, E.M., Peixinho, J. M. Formation of cylindrical shells via sphere packing from fluidized beds. The European Physical Journal E, v. 49, 30, 2026. https://doi.org/10.1140/epje/s10189-026-00573-z}}}
	
	\author[1,2]{Vinícius P. S. Oliveira}
	\author[1,3]{Danilo S. Borges}
	\author[1]{Erick M. Franklin}
	\author[2]{Jorge Peixinho}
	
	\affil[1]{Faculdade de Engenharia Mec\^anica, UNICAMP-Universidade Estadual de Campinas, Rua Mendeleyev, 200 Campinas, SP, Brazil}
	\affil[2]{Laboratoire PIMM, CNRS, Arts et Métiers Institute of Technology, Cnam, \\
		151 boulevard de l’H\^{o}pital, Paris, France}
	\affil[3]{Faculty of Physics, University of Duisburg-Essen, 47057 Duisburg, Germany}
	
	\abstract{The results of a numerical investigation of fluidized beds of spherical particles in a narrow vertical cylindrical pipe, with particular attention to the spontaneous settling along the wall, are reported.
		Starting from a steady fluidized state, the particles fluctuate because of fluid-particle, particle-particle, and particle-wall interactions.
		The particles are heavier than the fluid, with diameters $d$ yielding ratios of pipe to particle diameters $D/d=4.3$ and $4.7$. 
		For given ranges of flow velocities and bed sizes, particles settle on the wall, with a decrease in the bed height and particle fluctuations. Either a glass- or crystal-like shell forms along the pipe wall, in qualitative agreement with previous experiments.
		The polydispersity and the particle-particle friction are varied to test the stability of the particulate shell formation.
		The shell structure is analyzed by unwrapping it in a plane and locating all particles and their contact points, and we find that it exhibits a hexagonal lattice with a defects density that increases with polydispersity. 
		The shell formation is hindered by polydispersity, and there exists a critical point for polydispersity above which a crystal-like shell is unstable. 
		In a particular case of bidisperse beds, the crystal-like shell only appears when the particle-particle friction is high enough. 
		Finally, we compute the contact forces within particle-particle chains and in particle-wall contacts, which sustain the cylindrical shell, highlighting the dominant role of particle-particle forces.}
	
	
	
	\maketitle
	
	\section{Introduction}
	
	Solid particles confined in a vertical tube in the presence of an ascending fluid can sink, if their weight is higher than the forces caused by the fluid, or rise if the contrary is true. 
	They can also remain suspended in approximately the same place if the weight of particles equals the fluid forces, this latter case being known as fluidized bed. In fluidized beds, part of the fluid energy is passed to the solid particles, and is eventually dissipated through inelastic particle-particle and particle-wall collisions or friction (depending on the regime of particle motion \cite{Andreotti2013}).
	As the ratio between the diameters of pipe $D$ and particles $d$ increases, a network of chains of contact forces develops, percolating the forces within the particles to the tube wall, and leading to the organization of particles in static structures. 
	One of these structures appears when particles settle and form an annular packing in the shape of a cylindrical shell, having been observed for very narrow beds within $4 \leq D/d \leq 10$ \cite{Cunez2020}.
	This packing is related to the crystallization of monodisperse hard spheres in cylinders \cite{Royall2024}, where a rich set of structural and thermodynamically stable ordered configurations are observed: zigzag and helical chains for small confinement, that is, when the cylinder diameter less than 4 times the particle diameter.
	The associated morphological richness was first studied by Pickett et al. \cite{Pickett2000}, who found that chiral order spontaneously develops for certain diameter ratios. 
	Mughal et al. \cite{Mughal2011,Mughal2012} later adopted a phyllotatic description to understand how the corresponding densest-packed structures at $2 < D/d < 2.71486$ arise, in which cases all spheres of any densest-packed structure are in contact with the cylindrical wall.
	Beyond this diameter regime, not all spheres touch the cylinder wall, which eventually results in a separation between core and shell particles \cite{Fu2016}. 
	Exotic arrangements, complex helices, and limit periodic structures follow \cite{Lee2017,Kurban2025}. 
	More or less systematic numerical exploration of these structures ended at $D/d\approx4$, but it is conceivable that larger diameters might accommodate even more unusual structures \cite{Mughal2012,Fu2017}.
	
	This spontaneous self-assembly of monodisperse particles into crystal-like structures were also observed in colloidal suspensions in capillary tubes \cite{Moon2004} for manufacturing microporous cylinders.
	Still, the spontaneous crystal-like shell has been observed in large scale configurations (without considering Brownian motion) \cite{Oliveira2023}, but the roles of friction and polydispersity are not yet fully understood.
	
	A related old and popular problem in granular clogging is the discharge of beads from a vertical cylinder through a pierced disc \cite{Beverloo1961,Zuriguel2005}. 
	Many researchers have studied this confinement instability, both using experiments and numerical simulations.
	2D (two-dimensional) DEM (discrete element method) simulations have shown that the probability of clogging is correlated with the friction coefficient \cite{Zhou2025}.
	For larger $D/d$, 3D (three-dimensional) DEM simulation have been carried out \cite{Pournin2005,Pournin2007,Zhang2021}, showing that low friction favors the formation of particle rings around the cylinder wall, being an important element for the shell stability.
	
	This paper presents a numerical investigation on the spontaneous crystallization of narrow beds consisting of either mono or polidisperse grains that were initially fluidized. 
	We are particularly interested in the formation of a static structure in the shape of a cylindrical shell. For that, we carried out CFD-DEM (computational fluid dynamics - discrete element method) simulations in which we varied the particle diameter, the polydispersity, and the particle-particle friction. 
	We find that numerical simulations reproduce previous experiments in narrow beds, that polydispersity hinders the formation of the crystal-like shell, and that in the case of monodisperse beds they only appear for high coefficients of solid friction. 
	In addition, we compute the contact forces within particle-particle chains and in particle-wall contacts, which sustain the cylindrical shell, and show that most of the load is sustained by the lateral wall in comparison with the bottom of the tube (bottom boundary). 
	Our results bring new insights into both the spontaneous defluidization of fluidized beds and the formation of crystal-like packings in the form of cylindrical shells.
	
	\section{Methodology}
	
	Eulerian–Lagrangian simulations were performed using the \texttt{OpenFOAM} CFD open-source code for the continuous phase and the \texttt{LIGGGHTS} DEM also an open-source code for the discrete phase. 
	Both solvers were coupled using the unresolved coupling scheme available in the \texttt{CFDEM} framework \cite{Kloss2012,Goniva2012}.
	
	The DEM part computes in a Lagrangian framework the linear (Eq. \ref{Fp}) and angular (Eq. \ref{Tp}) momentum conservation for each solid particle,
	
	\begin{equation}
		m_{p}\frac{d\vec{u}_{p}}{dt}= \vec{F}_{d} + \vec{F}_{press}+ \vec{F}_{\tau} + \vec{F}_{am} + m_p\vec{g} + \vec{F}_{c} \,\,,
		\label{Fp}
	\end{equation}
	
	\begin{equation}
		I_{p}\frac{d\vec{\omega}_{p}}{dt}=\vec{T}_{c} \,\,,
		\label{Tp}
	\end{equation}
	
	\noindent where, for each solid particle, $m_{p}$ is the mass, $\vec{u}_{p}$ is the velocity, $I_{p}$ is the moment of inertia, $\vec{\omega}_{p}$ is the angular velocity, $\vec{F}_{c}$ is the resultant of contact forces between solids, $\vec{T}_{c}$ is the resultant of contact torques between solids, $\vec{F}_{d}$ is the drag force caused by the fluid on particles, $\vec{g}$ is the gravitational acceleration, $\vec{F}_{press}$ is the force caused by the fluid pressure, $\vec{F}_{\tau}$ is the force caused by the deviatoric stress tensor, and $\vec{F}_{am}$ is the added mass force. The equations for each of these forces are available in Appendices A and B. 
	We neglect torques caused directly by the fluid because those due to contacts are much higher \cite{Tsuji,Tsuji2,Liu}.
	
	The CFD part computes in an Eulerian framework the mass (Eq. \ref{mass}) and (Eq. \ref{qdm}) momentum equations of conservation,
	
	\begin{equation}
		{\frac{\partial{\rho_{f}\varepsilon_{f}}}{\partial{t}}+\nabla\cdot(\rho_{f}\varepsilon_{f}\vec{u}_{f})=0} \,\,,
		\label{mass}
	\end{equation}
	
	\begin{dmath}
		\frac{\partial{\rho_{f}\varepsilon_{f}\vec{u}_{f}}}{\partial{t}} + \nabla \cdot (\rho_{f}\varepsilon_{f}\vec{u}_{f}\vec{u}_{f}) = \alpha_{f} \rho_{f} \vec{g} -\varepsilon_{f}\nabla P + \varepsilon_{f}\nabla\cdot \vec{\vec{\tau}}_{f} - \frac{\vec{F}_{exch}}{V_{cell}} \,\,,
		\label{qdm}
	\end{dmath}
	
	\noindent where $\vec{u}_{f}$ is the mean velocity and $\varepsilon_{f}$ the volume fraction of the fluid phase, $V_{cell}$ is the volume of the considered cell, $P$ is the fluid pressure, $\vec{\vec{\tau}}$ is the deviatoric stress tensor of the fluid, and $\vec{F}_{exch}=\vec{F}_{D} + \vec{F}_{am}$ is an exchange term (the forces due to the pressure gradient and deviatoric stress tensor are split from the remaining fluid-particle forces when obtaining the volume-averaged Eqs. \ref{mass} and \ref{qdm}). 
	Eqs.~\ref{mass} and ~\ref{qdm} are solved using the PISO (pressure-implicit with split operators) algorithm.
	
	The simulation domain was a vertical cylindrical column with diameter $D=25.4$ mm and a total height of 450 mm.
	Several mesh configurations were tested before reaching validation with experimental data, as summarized in Table~\ref{tab:val}. 
	The best match with the experimental validation case was obtained with mesh E, yielding a relative error of $\approx 1$\%, which was considered acceptable.
	This mesh leads to a resolution of 2-3 grid points to a particle diameter.
	The bottom cross-sectional area of the cylinder was defined as an inlet with a constant vertical bulk velocity, $U$. 
	The top boundary was defined as an outlet with the same bulk velocity, $U$, and zero pressure.
	
	\begin{table}[ht]
		\centering
		\begin{tabularx}{\columnwidth}{
				>{\centering\arraybackslash}X 
				>{\centering\arraybackslash}m{1.8cm} 
				>{\centering\arraybackslash}m{1cm} 
				>{\centering\arraybackslash}X 
				>{\centering\arraybackslash}m{1cm}}
			\hline
			\rule{0pt}{2.4ex}\multirow{2}{*}{Mesh} & \multirow{2}{*}{No. of cells}  & \multirow{2}{*}{\parbox{\linewidth}{$\bar{V}_{cell}$ \\ (mm$^3$)}} & $H$ & $H_{error}$  \\
			& & & (cm) & (\%) \\
			\hline
			A & 9 600 & 14 & 4.72 & 2.9 \\
			B & 16 530 & 7.7 & 4.72 & 3 \\
			C & 20 625 & 5.9 & 4.65 & 8 \\
			D & 22 275 & 5.5 & 4.70 & 6 \\
			E & 22 605 & 5.4 & 4.80 & 1 \\
			\hline
		\end{tabularx}
		\caption{Mesh sensitivity study comparing the number of elements, the average cell volume, $\bar{V}_{cell}$, the bed height $H$, and the associated error relative to experimental data.}
		\label{tab:val}
	\end{table}
	
	The simulation procedure consisted of three main steps: (i) the particles were randomly placed in the cylinder and allowed to settle under gravity, (ii) a vertical fluid velocity ramp was applied over 0.8 s until reaching $U=13.2$ cm/s and (iii) the fluid flow was maintained constant for a simulation time of 60 s.
	The Reynolds number is about 3350, which corresponds to a transition regime.
	Thus, the effect of turbulence is taken into account using a k-epsilon ($k$-$\epsilon$) model.
	All computations were carried out using a desktop computer with 8 processors. 
	
	The setup was adapted from a previous numerical configuration \cite{Cunez2020b}, and considering the experimental data-set generated by \cite{Oliveira2023}. 
	The simulation parameters of what we call in the following validation case are listed in Table~\ref{tab:parameters}. 
	The number of particles, $N$, and the bulk velocity, $U$, were systematically varied. 
	
	\begin{table*}[ht]
		\centering
		\begin{tabular}{lcc}
			\hline
			Parameter & Symbol (Unit) & Value \\
			\hline
			Average diameter & $d$ (mm) & 5.88  \\
			Dispersity & $\sigma/d$ & 0.0022 \\
			Bulk velocity & $U$ (cm/s) & 13.2 \\
			Number of particles & $N$ & 100 \\
			Particle density & $\rho_p$ (kg/m$^3$) & 2330 \\
			Fluid density & $\rho_f$ (kg/m$^3$) & 1000 \\
			Dynamic viscosity & $\eta$ (Pa.s) & 0.001 \\ 
			Particle-particle coefficient of friction & $\mu_{p-p}$ & 0.05 \\
			Wall-particle coefficient of friction & $\mu_{p-w}$ & 0.1 \\
			Particle-particle coefficient of restitution & $e_{p-p}$ & 0.1 \\
			Particle-wall coefficient of restitution & $e_{p-w}$ & 0.1 \\
			\hline
		\end{tabular}
		\caption{Simulation parameters used in the reference case}
		\label{tab:parameters}
	\end{table*}
	
	We note that the validation case corresponds to experiments reported in Ref. \cite{Oliveira2025} for PTFE particles with a small but measurable dispersity. 
	In the present simulations, the small polydispersity in the experiments is approximated by a tridisperse distribution of diameters: 25\% of particles with a diameter of 5.86 mm, 50\% of 5.88 mm and 25\% of 5.90 mm. 
	The mean absolute deviation, $\sigma=(0.02+0.02)/3$ mm, leads to the relative dispersity coefficient $\sigma/d=0.0022$, as presented in Table \ref{tab:parameters}.
	In addition, the particle–particle friction coefficient $\mu_{p-p}$ and the particle size polydispersity $\sigma$, represented here as relative bidispersity, $\sigma/d$, were varied to investigate their influence on the formation of crystal-like structures via a structure index, $\chi$, where 1 represents the fluidized state and 0 represents the crystal-like shell.
	
\begin{figure*}[ht]
		\centering
		\includegraphics[width=0.8\linewidth]{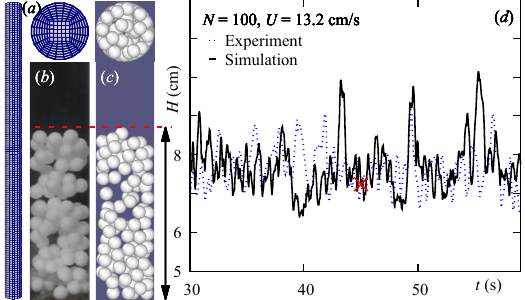}
		\caption{Validation case. (a) Snapshot of an experimental fluidized bed ($N=100$ PTFE particles, $d=5.88\pm0.01$ mm and $U=13.2$ cm/s), (b) snapshot at time-step $t=45$ s from the simulation (see parameters in Table \ref{tab:val}) together with the top view and (c) comparison of bed height time evolution, $H(t)$, for experiments and simulations, where the red cross indicated $t=45$ s.}
		\label{fig:val}
	\end{figure*}

	Figure \ref{fig:val} presents the mesh E and the results of the simulations of the validation case with the parameters from Table \ref{tab:parameters}.
	The parameters were the same as in the experiments, including the diameter, the particle-particle (PTFE-PTFE), $\mu_{p-p}$, and the particle-wall (PMMA-PTFE), $\mu_{p-w}$, friction coefficients, that were measured \cite{Oliveira2025}.
	The coefficient of restitution was fixed to 0.1.
	The comparison of the bed height $H(t)$ between the experiments and simulations is in good agreement both in terms of mean and fluctuations.
	
	The numerical simulation returns the instantaneous positions of all particles, being a good tool to identify the formation of the cylindrical packing along the wall.
	In the analysis that follows, specifically when the particles form a crystal-like structure along the wall with particles centers on a cylindrical plane at a radius position $r\simeq(D-d)/2$ from the pipe axis, the coordinates of particle positions are then unwrapped.
	Voronoi tessellation was used to quantify the organization of particle arrangements in the 2D domain. 
	In this method, the space is divided into polygonal cells, each associated with a single particle, such that any point within a cell is closer to its associated particle than to any other. 
	The edges of the Voronoi polygons correspond to the boundaries between neighboring particles. 
	From each triplet of adjacent particles, we computed the angles between nearest neighbors, and from the tessellation we determined the coordination number, i.e., the number of nearest neighbors for each particle.
	In addition, the unwrapped diagrams of the contact forces represent the force distribution along the cylindrical shell, designating the force chains and arches.
	
	\section{Results}
	
	Our numerical results reproduce qualitatively the shell formation and particle arrangement, while quantitatively reproducing the bed expansion observed in experiments under comparable operating conditions \cite{Oliveira2025}.
	However, different from experiments, we can impose larger variations of the pertinent parameters, and have access to all particle positions, solid-solid contacts, and forces. 
	In this section, we show and analyze the appearing structures in both macro (bed) and micro (particles and contacts) scales.
	
	\subsection{Crystal-like structures as functions of $U$ and $N$}
	
\begin{figure}[ht]
		\centering
		\includegraphics[width=1.0\linewidth]{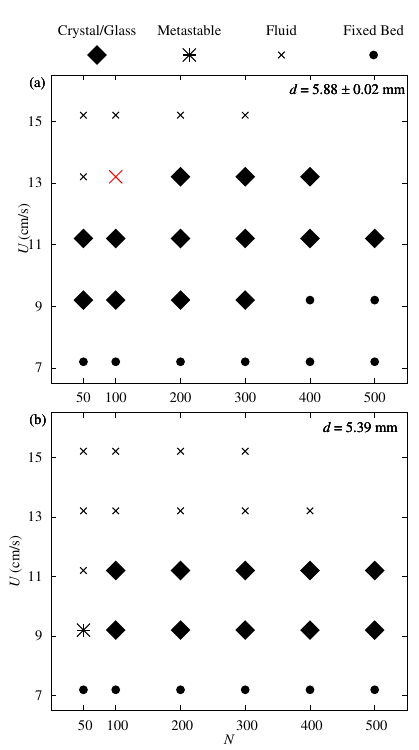}
		\caption{Regime map in the $N-U$ space, indicating the fixed bed, the fluidized, the metastable and the crystal-like structures (a) for $d=5.88\pm0.02$ mm (tridisperse, $D/d \simeq 4.3$) and (b) for $d=5.39$ mm (monodisperse, $D/d \simeq 4.7$). 
			The (fluidized) validation case is marked using a large red cross.}
		\label{fig:diag}
	\end{figure}

	The occurrence of the crystal-like structure, within the maximum real time of 60 s, is plotted in a $U$ versus $N$ map in Fig. \ref{fig:diag}(a), for a tridisperse bed and $D/d \simeq 4.3$, where $d$ is the average particle diameter.
	According to the map, the bed remains static at low values of $U$, as expected since the fluid velocity is not high enough to suspend part of the particles (it is the so-called minimum fluidization velocity $U_{mf}$). 
	For higher values of $U$, the fluid flow is able to sustain the weight of particles and move them around, and, if the values of $U$ are moderate, the particles eventually move towards the wall and pile-up in a self-organized fashion.
	Under these conditions, the core of the bed becomes hollow and the particle motion ceases, in what we call a crystal-like structure. 
	Finally, at large values of $U$, the structure becomes unstable and the particles are entrained further downstream by the flow.
	
	\subsection{Effect of \textit{d}}
	
	Figure~\ref{fig:diag}(b) presents the $U\times N$ map illustrating the observed regimes for a smaller particle diameter and monodisperse bed and $D/d \simeq 4.7$.
	Again, the formation of different structures depends primarily on $U$ and $N$: when $U<U_{mf}$, the bed remains static, when $U$ assumes moderate values the bed fluidizes but eventually undergoes a transition where particles migrate to the wall, and for higher values of $U$ the bed fluidizes. 
	In some cases, a fluidized bed alternates with the static structure, and we identify those cases as metastable \cite{Oliveira2023,Oliveira2025}. 
	In the present simulations, this appeared in the monodisperse bed for the smallest $U>U_{mf}$ and smallest $N$.
	We note that the condition for the occurrence of crystal-like structures are slightly different when the particle diameter decreases.
	More specifically, when $d=5.88\pm0.02$ mm ($D/d\simeq4.3$) the system forms a cylindrical shell for $N=200$ to 400 at $U=13.2$ cm/s, whereas it fluidizes when $d=5.39$ mm ($D/d\simeq4.7$).
	We note also that the simulations are limited by the computational domain and simulations at larger $N$ and $U$ lead to particles escaping the domain.
	This is the reason why data at larger $N$ and $U$ are not available.
	
	Figure~\ref{fig:diag}(b) also shows the effect of particle size on the emergence of these patterns for monodisperse systems ($\sigma/d=0$). 
	Two different particle diameters were analyzed. 
	We observed that larger particles tend to favor the crystallized regime, while smaller particles are more likely to remain fluidized under the same flow conditions. 
	Interestingly, one of the cases that crystallized in the polydisperse scenario did not crystallize when made monodisperse.
	
	\subsection{Arrangement of particles within the shell}
	
\begin{figure}[ht]
		\centering
		\includegraphics[width=\linewidth]{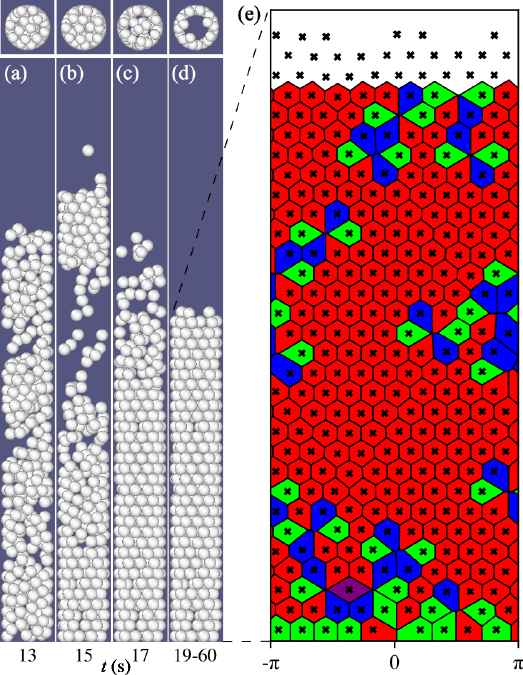}
		\caption{Snapshots (side and top views) and unwrapped crystal-like structure: $d=5.88\pm0.02$, $D/d \simeq 4.3$, $\sigma/d=0.0022$  (tridisperse) and $N=300$.
			(a-c) Fluidized bed at $t=13$, 15 and  17 s. 
			(d) Crystal-like structure that remains essentially fixed for $19\leq t \leq 60$ s.
			(e) Magnified unwrapped Voronoi analyses of the crystal-like structure where the black crosses represent the center of the particle and the color of the cell indicates the coordination number: red is 6 neighbors, green is 5 and blue is 7 neighbors.}
		\label{fig:patterns}
	\end{figure}

	Let's consider a case with a relatively large $N$ that forms the cylindrical shell, e.g. $d=5.88\pm0.02$ mm tridisperse ($D/d \simeq 4.3$ and $\sigma/d=0.0022$) and $N=300$.
	In Fig. \ref{fig:patterns}(a-d), snapshots of a sequence show that the initial fluidized bed starts to form the crystal-like structure from the bottom. 
	We note from the top view that the structure is hollow, forming a crystal-like cylindrical shell, whose principal feature is a hexagonal packing with defects that emerge.
	The cylindrical shell or crystal-like structure can be better examined through unwrapping of the surface at $r\simeq(D-d)/2$ and performing the Voronoi analysis together with periodic boundary conditions in the azimutal direction and special conditions at the top and bottom boundaries.
	The analysis highlights a regular/periodic distribution of the centers of particles with a limited number of defects corresponding to voids.
	For the majority of particles, the coordination number or number of neighbors is 6 (red cells), suggesting the role of the hexagonal lattice, known to be the optimal packing configuration.
	At the location of defects, the analysis specify particles with 5 (green cells) and 7 neighbors (blue cells), which can be verified considering the tip of the cells.
	
	\subsection{Effect of bidispersity}
	
\begin{figure}[ht]
		\centering
		\includegraphics[width=1.0\linewidth]{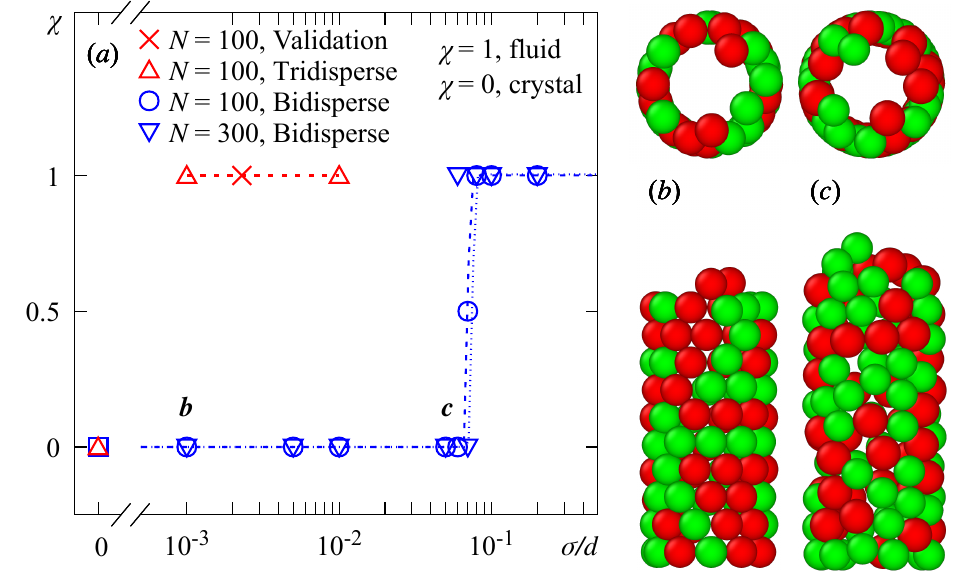}
		\caption{Shell packing state for bidisperse mixtures with $D/d \simeq 4.3$, $d=5.88$ mm and $U=13.2$ cm/s. 
			(a) Final state parameter, $\chi$, as a function of the relative particle diameter dispersion, $\sigma/d$. 
			The validation case is marked by a red cross. 
			Snapshots of crystal/glass structures for (b) $\sigma/d=0.001$ and (c), $\sigma/d=0.05$, where the small beads are colored green and the large red.}
		\label{fig:bidispersity}
	\end{figure}

	Polydispersity is often reported to hinder particle organization and suppress crystallization. 
	In this study, polydispersity was represented by a bidisperse set of particles, with two distinct diameters defined as $d=d\,(1\pm\sigma)$, where $d$ is the mean particle diameter and $\sigma$ is the deviation. 
	The relative bidispersity $\sigma/d$ was systematically varied from 0 (mondisperse) to 0.2. 
	Indeed, it is found that increasing $\sigma$ favors the onset of fluidized beds transition between fluidized and crystal-like structures taking place at $\sigma/d=0.09$ (see Fig. \ref{fig:bidispersity}).
	Comparing cases with different $N$ and same bidispersity, we observe similar patterns. 
	This suggests that $N$ does not influence defluidization within the ranges of our study.  
	Figure ~\ref{fig:bidispersity}(b–c) shows snapshots of crystal-like structures with $\sigma/d=0.001$ and 0.05, respectively, from which we notice that a more disordered structures for large $\sigma$. 
	Zhang et al. \cite{Zhang2021} proposed a shell growth mechanism suggesting granular segregation \cite{Zuriguel2006,Gray2018,Vo2021}; however, no particle segregation was clearly observed in the present simulations.
	The particle arrangement of the cylindrical shell for low $\sigma$ has hexagonal packing and helical particle alignments whereas particle with relatively large $\sigma$ show a more disordered packing arrangement.
	It is worth reporting that tridisperse mixtures remain fluidized from low $\sigma$.
	
	\subsection{Effect of friction}
	
\begin{figure}[ht]
		\centering
		\includegraphics[width=1.0\linewidth]{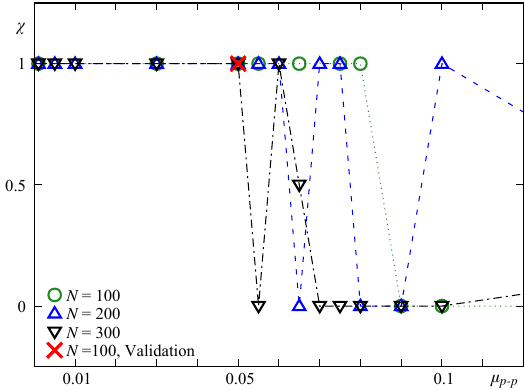}
		\caption{State parameter, $\chi$, of bidisperse mixtures with $D/d \simeq 4.3$, $d=5.88\pm0.02$ mm, $\sigma/d=0.07$ and $U=13.2$ cm/s, as function of the particle-particle coefficient of friction, $\mu_{p-p}$ for $N=100$, 200 and 300.
			The validation case (tridisperse: $\sigma/d=0.0022$ and $\mu_{p-p}=0.05$) is marked as a red cross.}
		\label{fig:friction}
	\end{figure}

	Numerical simulations allow the possibility to vary friction, both particle-particle friction, $\mu_{p-p}$, and particle-wall friction, $\mu_{p-w}$.
	We systematically analyzed the effect of $\mu_{p-p}$, varying it from 0.01 to 1, with a fixed relative bidispersity of $\sigma/d=0.1$. 
	For $N=100$, a transition from fluidized to crystal-like structure was observed at $\mu_{p-p} = 0.055$ (see Fig. \ref{fig:friction}). 
	These results indicate that the observed patterns depend, at least, on the interparticle friction, the polydispersity, and the number of particles. 
	Time seems to be an important factor in crystallized cases, metastability may be masked within the 60 s simulation window. 
	This suggests that crystallization may have a characteristic timescale that depends on both friction and superficial velocity, as reported in previous works \cite{Oliveira2023,Cunez2020}.
	
	\subsection{Contact forces}
	
\begin{figure*}[h]
		\centering
		\includegraphics[width=1.0\linewidth]{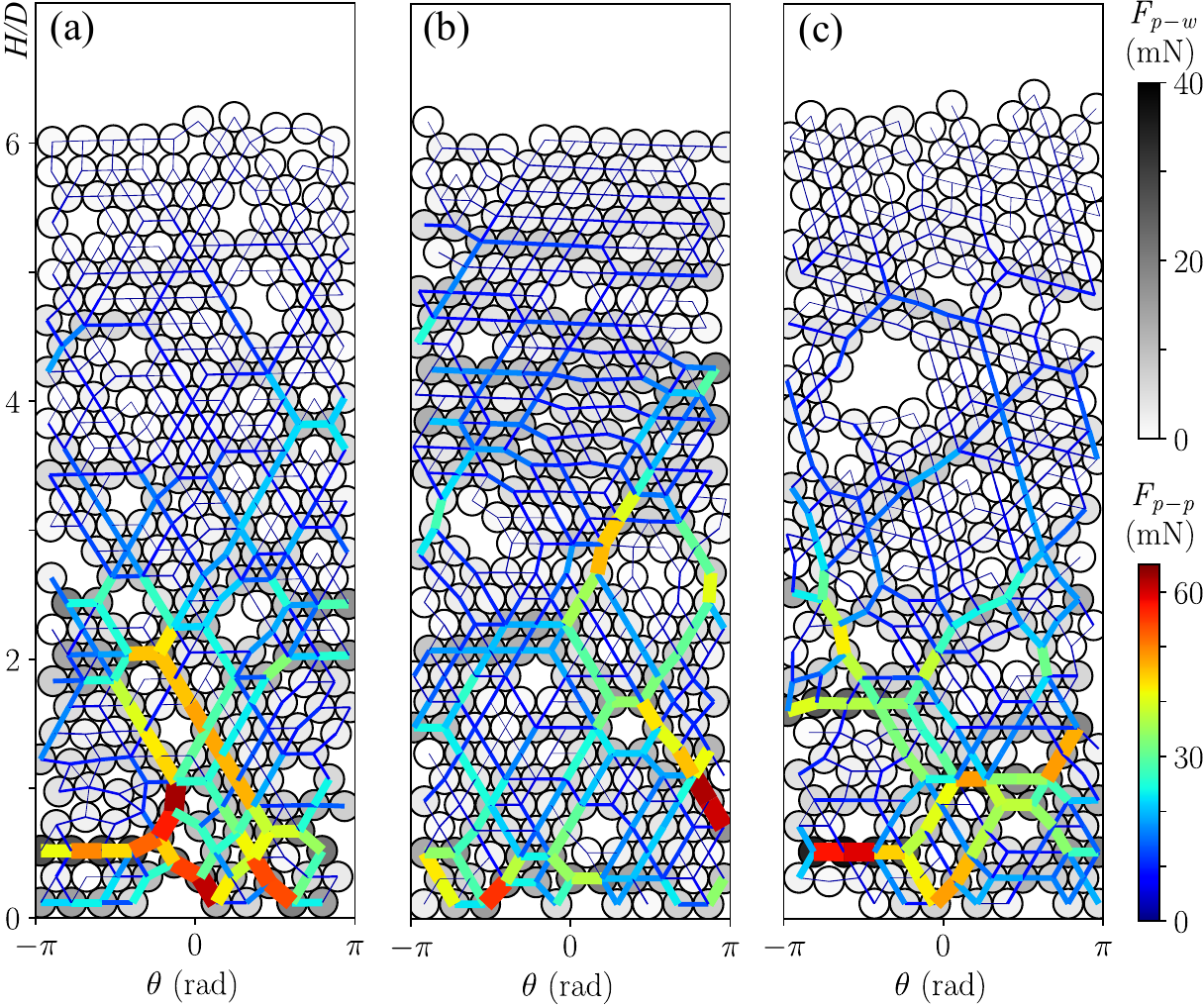}
		\caption{Contact force diagrams for different tridisperse cases with $D/d \simeq 4.3$, $d=5.88\pm0.02$ mm, $N=300$ and $U=13.2$ cm/s. 
			The intensity of force between particles $F_{p-p}$ is represented by colored chains and the intensity of force between particles and the wall is represented as a gray-scale particle color. 
			(a) $\sigma/d=0.0022$ and $\mu_{p-p}=0.05$, 
			(b) $\sigma/d=0.0022$ and $\mu_{p-p}=0.1$ and 
			(c) $\sigma/d=0.0011$ and $\mu_{p-p}=0.1$}
		\label{fig:contact}
	\end{figure*}

	In CFD-DEM simulations, it is possible to compute and store the forces acting on each individual particle, so that we have access to the distributions of the different forces within the cylindrical shell. 
	For example, in Fig. \ref{fig:contact}, the particle-particle (color lines of varying thickness) and particle-wall (in gray color) contact forces are displayed allowing to visualize the force arches propagating horizontally and obliquely through the shell.
	Three cases have been selected to exemplify the effect of $\sigma$ and $\mu_{p-p}$ on the bed structure. 
	We first note that the defects (or voids) are surrounded by relatively strong force chains. 
	Second, we notice that the levels of forces are relatively similar between the different cases, with the structure being mainly sustained by the particle-particle forces, $F_{p-p}$, applying forces to the cylindrical wall mostly in the region near the bottom ($H/D\leq2$).
	
	In order to quantify the relative roles of the lateral wall and bottom boundary on supporting the shell structure, we computed the total forces exerted on particles by the lateral wall, $\vec{F}_{w}$, and bottom boundary, $\vec{F}_{b}$. 
	In addition, we decomposed them into their tangential (parallel to the considered boundary) and normal components, given by subscripts $t$ and $n$, respectively, as shown in Eqs. \ref{eq:wall_forces} and \ref{eq:bottom_forces}: 
	\begin{equation}
		\vec{F}_{w} = \sum^{N}_{i=1}{\vec{F}_{i,p-w}} = \vec{F}_{t,w} + \vec{F}_{n,w}
		\label{eq:wall_forces}
	\end{equation}
	
	\begin{equation}
		\vec{F}_{b} = \sum^{B}_{j=1}{\vec{F}_{j,p-b}} = \vec{F}_{t,b} + \vec{F}_{z,b}
		\label{eq:bottom_forces}
	\end{equation}
	Then, we computed the vertical component of the tangential component of the wall forces, $F_{z,w}$.
	The values of $F_{n,w}$, $F_{t,w}$, $F_{z,w}$, and $F_{z,b}$ are reported in Tab. \ref{tab:F} for the cases shown in Fig. \ref{fig:contact}. 
	The standard deviation of $F_{z,w}$ and $F_{z,b}$ are time-averaged over a period of 2 seconds indicating that force chains form and their force intensity fluctuates over time.
	We also note that approximately half of the relative weight of particles (approximately 417 mN) is supported by the bottom, while only a small proportion is supported by the lateral wall (the remaining of the load is supported by the fluid flow).
	This means that the lateral wall only supports few percents of the load supported by the bottom boundary. 
	This is different from particle filled tubes, for which a strong redirection of forces takes place and Janssen effect is expected \cite{Andreotti2013}. 
	In the case of cylindrical shells, the load percolates to the bottom boundary, as shown, indeed, by the network of contact forces presented in Fig. \ref{fig:contact}.
	
	\begin{table}[h]
		\centering
		\begin{tabularx}{\columnwidth}{
				>{\centering\arraybackslash}m{0.5cm} C C C C
				>{\centering\arraybackslash}m{0.8cm}
				>{\centering\arraybackslash}m{0.9cm}
			}
			\hline
			Fig. & $\sigma/d$ & $\mu_{p-p}$ & $F_{n,w}$ & $F_{t,w}$ & $F_{z,w}$ & $F_{z,b}$ \\
			& & & (mN) & (mN) & (mN) & (mN) \\
			\hline
			6(a) & 0.002 & 0.05 & 1472 & 116 & $8\pm 21$ & $205\pm2$ \\
			6(b) & 0.002 & 0.10 & 1562 & 230 & $7\pm 33$ & $218\pm2$ \\
			6(c) & 0.001 & 0.10 & 1188 & 194 & $6\pm 27$ & $187\pm3$ \\
			\hline
		\end{tabularx}
		\caption{Summary of the forces for the crystal-like shell cases varying $\sigma/d$ and $\mu_{p-p}$, presented in Fig. \ref{fig:contact} (tridisperse with $D/d \simeq 4.3$, $d=5.88\pm0.02$ mm, $N=300$ and $U=13.2$ cm/s).
			Summation of the particles forces on the cylindrical wall, the normal, $F_{n,w}$, the tangential, $F_{t,w}$, and the axial component $F_{z,w}$.
			$F_{z,b}$ is the total force exerted by the particles on the bottom or inlet.}
		\label{tab:F}
	\end{table}
	
	\section{Conclusions}
	
	In conclusion, CFD-DEM simulations of fluidized beds have been performed in order to better understand the formation of spontaneous shell crystal-like structures, occurring for relatively monodispersed particles under moderate upwards flow in a confined cylinder. Our simulations were carried out for ratios of pipe to particle diameters $D/d=4.3$ and $4.7$ in ensembles of $N=100$ to 500 particles.
	The simulations, although limited in confinement and size domains, reproduce qualitatively some features of the cylindrical shell: the dependence on $U$ and $N$.
	By varying the bidispersity, $\sigma$, and the friction, $\mu_{p-p}$, the simulations suggest critical behavior, that is a threshold $\sigma$ that renders shell unsustainable and a critical $\mu_{p-p}$ above which the shell is easily obtained. 
	Our results also show that the lateral wall only supports few percents of the vertical load supported by the bottom boundary, indicating that redirection of forces is much lower in the cylindrical shell with respect to a filled tube. 
	In the future, more complex cases, such as spherocylinders, can be considered.
	
	\backmatter
	
	\bmhead{Acknowledgments}
	
	We acknowledge the support of FAPESP-CNRS Grant n\textdegree 2024-02440-2. EMF, VPSO and DSB are grateful for the support of FAPESP (Grant Nos. 2018/14981-7, 2020/00221-0, 2022/01758-3, 2024/02440-2, 2024/13295-3).
	
	\section*{Author contribution statement}
	
	VPSO performed the simulations, wrote the image analysis code and visualization codes, and edited the manuscript. 
	DS contributed to methodology, investigation, writing-review and editing, and supervision.
	EMF and JP contributed to conceptualization, funding acquisition, resources, methodology, formal analysis, investigation, writing-review and editing, and supervision.
	
	\bmhead{Conflict of interest}
	
	The authors have no relevant financial or non-financial interests to disclose.
	
	\bmhead{Data Availability Statement}
	
	The data that support this article are available at \url{https://doi.org/10.17632/hdh8bxhfmd.1} 
	
	\begin{appendices}
		
		\section*{Appendix A  Fluid-particle interaction forces}
		\label{appendix_fluidforces}
		
		In the computations, the pressure forces $\vec{F}_{press}$ are given by Eq. \ref{Fp_appendix}, forces $\vec{F}_{\tau}$ due to the deviatoric stress tensor by Eq. \ref{Ft_appendix}, the added mass force $\vec{F}_{am}$ by Eq. \ref{Fam_appendix}, and the drag forces $\vec{F}_d$ by Eq. \ref{Fd_appendix},
		
		\begin{equation}
			\vec{F}_{press} = -V_p\nabla P \,\,, \tag{A1} 
			\label{Fp_appendix}
		\end{equation}
		
		\begin{equation}
			\vec{F}_{\tau} = V_p \nabla\cdot\vec{\vec{\tau}} \,\,, \tag{A2} 
			\label{Ft_appendix}
		\end{equation}
		
		\begin{equation}
			\vec{F}_{am} = \frac{1}{2} V_p \rho_{f}\left(\frac{d\vec{u}_{f}}{dt}-\frac{d\vec{u}_{p}}{dt}\right) \,\,, \tag{A3} 
			\label{Fam_appendix}
		\end{equation}
		
		\begin{equation}
			\vec{F}_d = \sum_{i}^{n_{p}}\frac{V_{p}\beta_G}{1-\alpha_{f}}\left(\vec{u}_{p}-\vec{u}_{fp}\right) \,\,, \tag{A4} 
			\label{Fd_appendix}
		\end{equation}
		
		\noindent where  $V_{p}$ is the volume of each solid particle, $\vec{u}_{fp}$ is the fluid velocity at the particle position, and $\beta_G$ is a model parameter (we considered the Gidaspow correlation for the drag force \cite{Gidaspow1994}).
		
		\section*{Appendix B Solid-solid contact forces}
		\label{appendix_contactforces}
		
		The solid-solid contact forces, $\vec{F}_{c}$, and torques, $\vec{T}_{c}$, are given by Eqs. \ref{Fc} and \ref{Tc},
		
		\begin{equation}
			\vec{F}_{c} = \sum_{i\neq j}^{N_c} \left(\vec{F}_{c,ij} \right) + \sum_{i}^{N_w} \left( \vec{F}_{c,iw} \right) \,\,,\tag{B1} 
			\label{Fc}
		\end{equation}
		
		\begin{equation}
			\vec{T}_{c} = \sum_{i\neq j}^{N_c} \vec{T}_{c,ij} + \sum_{i}^{N_w} \vec{T}_{c,iw} \,\,,\tag{B2} 
			\label{Tc}
		\end{equation}
		
		\noindent where $\vec{F}_{c,ij}$ and $\vec{F}_{c,iw}$ are the contact forces between particles $i$ and $j$ and between particle $i$ and the wall, respectively, $\vec{T}_{c,ij}$ is the torque due to the tangential component of the contact force between particles $i$ and $j$, and $\vec{T}_{c,iw}$ is the torque due to the tangential component of the contact force between particle $i$ and the wall. 
		$N_c-1$ is the number of particles in contact with particle $i$, and $N_w$ is the number of particles in contact with the wall. 
		In Eqs \ref{Fc} and \ref{Tc}, we apply the Hertz–Mindlin contact model.
		
		\section*{Appendix C Dealing with large particles in unresolved computations}
		\label{appendix_bigparticle}
		
		In the simulations, a method for called bigParticle is used for the fluid-particle interaction \cite{Lomholt2003,Abbas2007}, in which an artificial porosity is added to each solid particle while keeping the same mass (increasing, thus, their volume). 
		This is useful when particles are relatively large with respect to the CFD mesh, keeping the CFD code running in the unresolved case (which use volume averaged equations), so that a local void fraction is associated meshes occupied by particles  even when the particle is larger than the mesh size.
		
	\end{appendices}
	

\end{document}